\documentclass{article}
\usepackage{hyperref}
\usepackage{amsmath,amsthm,amssymb}
\usepackage{amsfonts}
\usepackage{booktabs}
\usepackage{graphicx}
\usepackage{soul}
\usepackage[export]{adjustbox}
\usepackage[boxed,noend]{algorithm2e}
\usepackage{float}
\usepackage{tikz}
\usepackage[all]{xy}
\usepackage{enumerate}
\usepackage{tkz-euclide}
\usetikzlibrary{calc}
\usetkzobj{all}

\newcommand{\cL}{\mathcal{L}}

\newcommand{\cS}{\mathcal{S}}

\newcommand{\bR}{\mathbb{R}}

\newcommand{\bZ}{\mathbb{Z}}

\def\vs{\vspace*}
\def\hs{\hspace*}
\def\lng{\langle}
\def\rng{\rangle}

\begin{document}
\title{Techniques in Lattice Basis Reduction}
\author{Bal K. Khadka and  Spyros  S. Magliveras  \\ \\
Department of Mathematical Sciences,\\ Florida Atlantic University \\ Boca Raton, FL 33431 \\ \\
email:  bkhadka@fau.edu,  spyros@fau.edu }
\date{}
\maketitle

\begin{abstract}
The credit on {\it reduction theory}  goes back to the work of Lagrange, Gauss, Hermite, Korkin, Zolotarev, and Minkowski. 
Modern reduction theory is voluminous and  includes the work of A. Lenstra, H. Lenstra and L. Lovasz who created the well known LLL algorithm, and many other researchers such as L. Babai and C. P. Schnorr who created significant new variants of basis reduction algorithms.
In this paper, we propose and investigate the efficacy of new optimization techniques to be used along with LLL algorithm. The techniques we have proposed are: i) {\it hill climbing (HC)}, ii) {\it lattice diffusion-sub lattice fusion (LDSF)}, 
 and iii) {\it multistage hybrid LDSF-HC}. The first technique relies on the sensitivity of LLL to permutations of the input basis $B$, and 
optimization ideas over the symmetric group $S_m$ viewed as a metric space.  The second technique relies on partitioning the
lattice into sublattices, performing basis reduction in the partition sublattice blocks, fusing the sublattices, and repeating.  
We also point out places where parallel computation can reduce run-times achieving
almost linear speedup. The multistage hybrid technique relies on the lattice diffusion and sublattice fusion and hill climbing algorithms.\\



\end{abstract}

\medskip

\pagestyle{plain}

\section{Prehistory}
The  development of  lattice basis reduction began in 1773 by L. Lagrange \cite{lagr1773}, C. F. Gauss \cite{gauss1801}, C. Hermite \cite{herm1850} (1850), 
A. Korkin and Y. R. Zolotarev \cite{korkin1873} (1873),  H. Minkowski\cite{minkowski1910} (1896), and B. L. van der Waerden \cite{warden1956} (1956) . \ The Hermite, Korkin, Zolotarev (HKZ \cite{stehle2007})  lattice basis reduction method is assumed to be the strongest but impractical reduction approach. 
In 1982, Arjen Lenstra,  Hendrik Lenstra and L\'{a}szl\'{o} Lov\'{a}sz \cite{lelelov1982},  
introduced a  polynomial time basis reduction algorithm to factor a rational polynomial, presently known as LLL, that  
approximates the {\em shortest vector} within a factor of $2^{n/2}$, where $n$ is the rank of the lattice. 
Thus, LLL is fast but weak due to the large approximation factor.  
In $1986,$ L. Babai \cite{babai1986}, proposed an algorithm that approximates the 
{\em closest vector} to a lattice point by a factor of $\big( \frac{3}{\sqrt{2}} \big)^n.$
 In $1987,$  C. P. Schnorr \cite{schnorr1987}  proposed a further refinement of the LLL algorithm 
that improved the shortest vector  approximation factor to $(1+\epsilon)^n.$ 
The {\em Block Korkin Zolotarev} (BKZ) algorithm was introduced and studied   \cite{babai1986, schnorr1992} by Babai and  Schnorr respectively.
  In 1993, M. Seysen \cite{seysen1993} showed that each full rank $n$ lattice $\cL$ has a basis $B$ and an associated basis $B^*$ of the reciprocal lattice $\cL^*$, with $S(B)\leq \exp(c_2\cdot (\ln n)^2)$, where $S(B)=\max_{1\leq i \leq n}(\|b_i\|\cdot \|b^*_i\|)$ and $c_2$ independent of $n$.
 This was an improvement on the relation  $S(B)\leq \exp(c_1\cdot n^{1/3})$ established by J. Hastard and J. C. Lagarias \cite{hastad1990}.   
 In 1994, Schnorr and Euchner \cite{schnorr1994} introduced  improved algorithms for BKZ and {\em deep insertion}, but the runtime of these algorithms  is not polynomial. 
\par
In 2001, M. Ajtai et al., \cite {ajtai2001} introduced a {\em Sieve} algorithm for the shortest lattice vector problem.  In $Eurocrypt \  2008,$ {\em N. Gama and P. Nguyen}
\cite{gama2008}  observed experimentally that  BKZ's practical runtime seems to grow exponentially with lattice dimension. 
In 2010, O. Dagdelen and M. Schneider, \cite{dagdelen2010}  presented a parallel version of the lattice enumeration algorithm using a multi-core CPU system. 

The shortest vector (SVP) and closest vector (CVP) problems, presently considered intractable, are algorithmic 
tasks that lie at the core of many number theoretic problems, integer programming, 
finding irreducible factors of polynomials, minimal polynomials of algebraic numbers, 
and simultaneous Diophantine approximation.  Lattice basis reduction also has deep and extensive connections with modern cryptography, and
cryptanalysis particularly in the post-quantum era. At present, lattice based cryptosystem do not appear to be vulnerable to quantum attacks. 
Also, the implementation of a lattice based cryptosystem is faster than the RSA or ECC system at an equivalent security level which is proposed in papers by T. G\"{u}neysu  et.al. in 2012,   \cite{guneysu2012} and by J. Howe in 2015, \cite{howe2015}.
 In 1982, A. Shamir \cite{shamir1982} proposed a polynomial-time algorithm for breaking the basic Merkle-Hellman cryptosystem.  
In 1985, J. C. Lagarias and A. M. Odlyzko, \cite{lagarias1985} solved low density knapsack problems.
In 1996, Hoffstein et al., \cite{hoffstein1998} proposed  NTRU: a ring theory based public key cryptosystem.
In 1997, D. Coppersmith \cite{copper1997} published a paper on small solutions to polynomial equations, and low exponent RSA vulnerabilities. 
In this paper, Coppersmith showed how to find small integer solutions to a modular polynomial in a single variable and extended the method to the multivariate case. 
In 1997, M. Ajtai and C. Dwork \cite{ajtai1997} designed a probabilistic public key cryptosystem whose security relies on the hardness of lattice problems. Their cryptosystem is quite impractical because of the massive data expansion, as it encrypts data bit-by-bit.

Inspired by the Ajtai-Dwork cryptosystem, O. Goldreich et. al., \cite{goldreich1997} proposed  a public key cryptosystem based on the closest vector problem in a lattice,  and J.-Y. Cai and T. W. Cusick, \cite{cai1999} introduced similar work on a lattice based public key cryptosystem in 1999. In 2001, C. Gentry, \cite{gentry2001} published a paper on Key Recovery and Message Attacks on NTRU-Composite, where he recovers  information about the private key used in NTRU. \ In 2001,  D. Coppersmith and A. Shamir, \cite{copper2001} proposed a new lattice basis reduction technique to cryptanalyze the NTRU scheme.
In 2009, Craig Gentry, \cite{gentry2009}, in his Ph.D. thesis, described a fully homomorphic encryption scheme based on ideal lattices.
Currently, there are many mathematicians vigorously contributing to lattice basis reduction theory, and improving best lattice bounds. 

\section{Preliminaries}
A lattice $\cL\subset \bR^{n}$, generated by $m$ linearly independent vectors $B=\{ b_{1},b_{2},...,b_{m\leq n} \} $ in $\bR^{n}$, is  the set of all integral linear combinations of the vectors of $B$. \[ \cL = \lng B \rng_{\bZ}=\{ a_{1} b_{1} + a_{2} b_{2} + \cdots + a_{m} b_{m} :  a_{i} \in \bZ \}. \]
Here, $B$ is called a {\it basis} for $\cL$, and $m$ the {\it rank} or {\it dimension} of $\cL$. For any two bases $B$ and $C$ of the lattice $\cL$, there is a unimodular matrix $U$ such that $B=UC$.
\subsection{Geometry of numbers}
Let $\mathcal{L}$ be a rank-$m$ lattice in $n$-dimensional Euclidean
space $\mathbb{R}^n$. The {\em first minimum} of the lattice, denoted $\lambda_{1}(\mathcal{ L}),$ is the length of a shortest nonzero vector $v_{1}\in\mathcal{ L}$.
The {\em second minimum} of the lattice, denoted by $\lambda_{2}(\mathcal{ L}),$ is the smallest real number $r$ such that there exist two $\bR$-linearly independent lattice vectors $v_{1}$ and $v_{2}$  such that $\|v_{1}\|, \|v_{2}\| \leq r.$ In general, for
$i = 1, 2, \ldots ,m,$ the $ i^{th}$  {\em successive minimum} of the lattice, denoted $\lambda_{i}\mathcal{( L)},$
is the smallest real number $ r$  such that there exist ``i"  $\bR$-linearly independent vectors
$v_{1}, v_{2},\ldots, v_{i} \in \mathcal{L}$ such that $\|v_{1}\|, \|v_{2}\|, \ldots , \|v_{i}\|\leq r.$
\subsection{Permutations and groups}
The symmetric group on $X = \{ 1, 2, \ldots , n \}$ is the collection of all  $n!$ permutations of the symbols of $X$, denoted by $ \cS_n$. $ \cS_n$ forms a group under the usual composition of functions. If $\pi \in \cS_n$, one of several ways of representing $\pi$ is as a two rows matrix of the form \ \large 
$\pi = (\begin{smallmatrix} 1 & 2 & \cdots & n \\ \pi(1) & \pi(2) & \cdots &\pi(n) \end{smallmatrix})$\normalsize, exhibiting the bijection.
\noindent
The \ {\it Cartesian form} \  of $\pi$ is simply the vector of images $[ \pi (1), \pi (2), \ldots , \pi(n)]$.

\noindent
The symmetric group $\cS_n$ can be made into a metric space by defining the distance between two permutations $x, y \in \cS_n$ as the number of places the Cartesian forms of $x$ and $ y$ differ.  Given a permutation $\pi \in \cS_n$ we associate with $\pi$ an $n \times n$ matrix $A=(a_{i,j})$, where \ $a_{i,j}=1$ \ if \ $\pi(i)=j$ and \ $a_{i,j}=0$ \ otherwise.
Such a matrix $A$ is called a permutation matrix. This is a matrix obtained from the identity matrix $I_n$ by permuting its columns by $\pi$.  We denote a single LLL operation on a permuted basis $B^{\pi}$ by $\Lambda(B^{\pi})$.

\noindent
We define the {\em radius} of a permutation as its {\em Hamming distance} from the identity permutation. A permutation $\pi_r$ of radius $r$ in the symmetric group $S_n$
 is said to be a {\em left permutation} if $r \leq \frac{n}{2}$,  otherwise, it is a {\em right permutation}.

\section{Sensitivity of  LLL under basis permutations}

A repeated application of LLL on the same ordered basis $B$ yields identical output. Consider a case where a single LLL algorithm can not completely reduce a lattice basis $B$. If we apply LLL lattice basis reduction after randomly (or systematically) reordering the original basis $B$, it will almost always yield a different basis.

Consider a basis $M_{54}$, an ideal lattice of rank 54, taken from the ideal lattice challange \cite{halloffame}. 
The length of the shortest vector ($l$)  after LLL on those lattice bases are given as \[l = 2745.73.\]
For each radius $r\in \{5, 10, \ldots , 50\}$, we select a random sample of $N=100$ permutations and compute $\Lambda(B^{\pi_i})$ for $1\leq i \leq 100$.
This yields a $100\times 1$ table (\ref{llloverpbasis})  of length of the shortest vector. Then we record the minimum value, maximum value, mean ($\mu$) and the standard deviation ($\sigma$) of this table. 

\begin{table}[H]
	\begin{center}
		\begin{tabular}{|c|c|c|c|c|c|}
			\hline  
			
			radius & $\min$ & $\max$  & $\mu$ & $\sigma$ & range \\
			\hline
			05 & 1946.87 & 2512.98   & 2241.26 & 115.51 & 566.11 \\
			\hline
			10 & 1968.17 & 2535.28 & 2256.89 & 102.84 & 567.11\\
			\hline
			15 & 1946.77 & 2532.94 & 2256.88 & 122.84 & 586.17\\
			\hline
			20 & 1946.87 & 2521.33 &2282.65 & 95.86 & 574.46\\
			\hline
			25 & 1946.87 & 2469.95 & 2264.28& 96.69 & 523.08\\
			\hline
			30 & 1968.17 & 2473.17 & 2263.23 & 101.65 & 505.00\\
			\hline
			35 & 1968.17 & 2558.02 & 2271.41 & 125.97 & 589.85\\
			\hline
			40 & 1946.77 & 2493.44 &2254.79 & 115.24 &  546.67\\
			\hline
			45 & 1968.17 & 2552.18 & 2284.52 & 109.11 & 584.01\\
			\hline
			50 & 2010.59 & 2490.98 & 2260.74 & 100.35 & 480.39\\
			\hline
		\end{tabular}
		\caption{LLL reduction over permuted basis}
		\label{llloverpbasis}
	\end{center}
\end{table}


\subsection{Effectiveness of right permutation}
Let $B$ be a lattice basis. Then, \ $\Lambda(B), \ \Lambda^2(B), \ldots ,\Lambda^n(B)$ \ are identical, that is $\Lambda$ is an idempotent operator. However,  table (\ref{llloverpbasis}) shows that the LLL algorithm is sensitive to permutations of the basis.
Based on these observations, we  analyze  the basis reduction pattern when  basis rows are permuted. 
In the experiments below, we take  a basis $B$ of an ideal lattice of rank 52 from
the {\it Hall of Fame} website \cite{halloffame} and reduce it by the LLL algorithm. Let $B^*=\Lambda(B)$.  $B^*$ is not of high quality and requires further reduction. We proceed as follows: for each possible radius  $r\in \{ 2,3,\ldots ,52 \}$, we  select a random sample $S_r=\{ \pi_{r,1}, \ldots, \pi_{r,N} \}$ of $N$ permutations, where $N\in \{100, 500 \}$. 
We compute \ $\Lambda({B^*}^{\pi})$ for all $\pi \in S=\bigcup\limits_{r=2}^{52} S_r$, \ and observe how often a basis gets reduced for each sample. 
The output parameters $l, L$, and $\log(wt)$  are the resulting lengths of the {\bf shortest} vector, {\bf longest} vector, and the $\log_{10}$ of the {\bf weight} of the reduced basis.
In the graph below, we use the following notation. \[  llb\leftarrow l_{\min}, \hs{5mm} lub\leftarrow L_{\min}, \hs{5mm}\text{and} \hs{5mm} mwt \leftarrow \min(\log_{10}wt) \]
where $\min$ represents the minimum of all sample for each radius.\\

\begin{figure}[H]
	\centering
	\includegraphics[width=.9\textwidth]{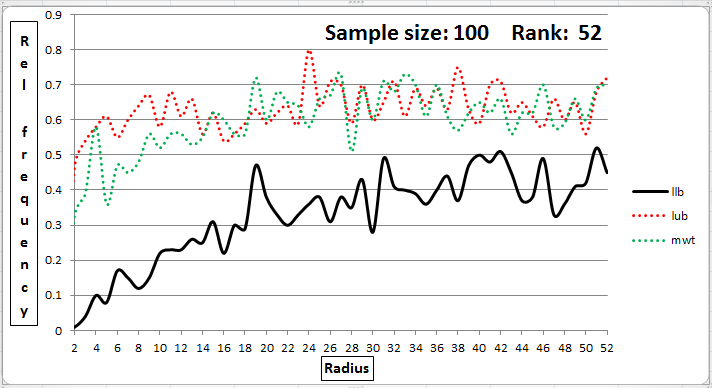}
\end{figure}
\vs{-2mm} 
\begin{figure}[H]
	\centering
	\includegraphics[width=.9\textwidth]{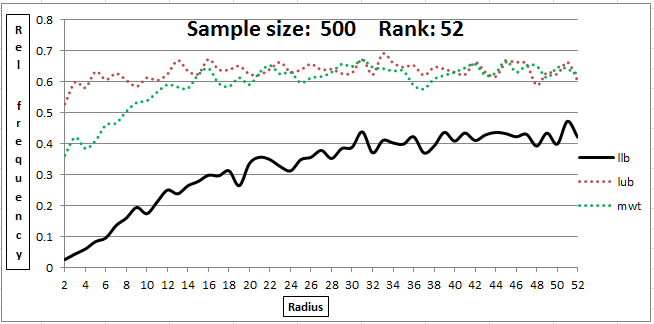}
	
\end{figure}
The relative frequency 0.6 corresponding to the radius $r$ in the graph below represents that the basis gets further reduced 60\% of the time if we rearrange the LLL reduced basis with a permutation of radius $r$. 
From above graphs, larger the permutation radius, larger the relative frequency for the shortest vector. 

Next, we compute the averages for $l, L, \log_{10}wt$ over the 500 permutations for each radius $r\in \{ 2,3,\ldots ,52 \}$ and analyze the pattern of these averages versus the radius. 

Further, we normalize the average values by using the transformation
\begin{center}normalized value $= \displaystyle \frac{\text{value - minimum value}}{\text{maximum value - minimum value}}$\end{center}
The normalized average values 0 and 1 corresponding to  the radii $r_1$ and $r_2$  represent the least and the greatest average among all experiments occurred under permutations of radii  $r_1$ and $r_2$ respectively.

\begin{figure}[H]
	\centering
	\label{navg}
	\includegraphics[width=.95\textwidth ]{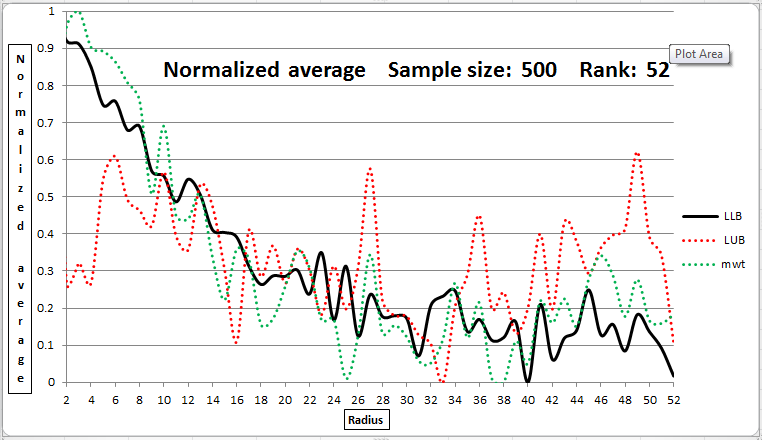}	
	
	\caption{Normalized average vs radius}
\end{figure}
In above experiments, the decreasing pattern of the curve corresponding to the shortest vector, makes clear that in average, the shortest vector length is achieved more efficiently if the radius of the permutation is larger. 
This means that the  length of the shortest vector is effectively reduced by an  application of permutations of  {\bf right} radii. This is true for number of experiments we have performed on lattice bases of different dimensions. Regarding those factors, we have proposed the hill climbing lattice basis reduction technique below.

\section{Hill Climbing Lattice Basis Reduction}
Hill climbing reduction is an optimization technique that attempts to search for a better solution locally and repeat the procedure until it finds a prescribed global solution. 
In our research, this method takes an advantage of the sensitivity of LLL to permuting the input basis. 
There are two types of Hill climbing lattice basis reduction techniques: 
\begin{enumerate}
	\item Reduction with a fixed radius (spherical random walk).
	\item Reduction with variable radius (spiral  random walk).
\end{enumerate}
These two methods are easy to understand from the description of the algorithms below. The second method differs from the first in the sense that the radii of successive permutations samples are systematically increased from one step to the next step, while radii are fixed in the first method.

\subsection{Hill climbing lattice basis reduction algorithm}
(Type I: {\em Fixed radius} )

\textbf{Input:} \ Basis \ $B_0=\{b_{1}, b_{2},\cdots ,b_{m\leq n}\}$ \ of \ $\cL \subset \bR^{n},$ and parameters: $k, \ p$.\\
\textbf{Output:} $v\in \cL , \, \  \text{where} \ \|v\|\leq m\cdot (\det \cL)^{1/m}$.

\begin{enumerate}	
	\item  $ i \leftarrow 1$, \ Compute $B_1 = \Lambda(B_0)$,
	\item Select a random sample of $k$ permutations of radius $r$, 
	$S_i = \{ \pi_{i,1},\ldots,\pi_{i,k} \} \subset \cS_m$, 
	\item Compute $\{ \Lambda(B_i^{\pi_{i,j }}) \ : \  1 \le j \le k \}$, and
	select $j$ such that $\Lambda(B_i^{\pi_{i,j}})$ achieves best reduction over the $i^{th}$
	step sample $S_i$,
	\item Update: $B_{i+1} \leftarrow \Lambda(B_i^{\pi_{i,j}})$ \ and let \ 
	$i \leftarrow i+1$,
	\item $\rightarrow 2$ \   if $\ (i \le p)$ \ and desired bound has not been achieved , {\bf else} end.
\end{enumerate}

(Type II: {\em Variable radius} )

\textbf{Input:} \ Basis \ $B=\{b_{1}, b_{2},\cdots ,b_{m\leq n}\}$ \ of \ 
$\cL \subset \bR^{n}$ \ and \ $r_0$, $rstep$, $k, p$, parameters.\\
\textbf{Output:} $v\in \cL , \, \  \text{where} \ \|v\|\leq m\cdot (\det \cL)^{1/m}$.

\begin{enumerate}	
	\item  $ i \leftarrow 1$, \ $r \leftarrow r_0$, \ Compute $B_1 = \Lambda(B_0)$,
	\item Select a random sample of $k$ right permutations of radius $r$, 
	$S = \{ \pi_{i,1},\ldots,\pi_{i,k} \} \subset \cS_m$, 
	\item Compute $\{ \Lambda(B_i^{\pi_{i,j }}) \ : \  1 \le j \le k \}$, and
	select $j$ such that $\Lambda(B_i^{\pi_{i,j}})$ achieves best reduction over $S$,
	\item Update: $B_{i+1} \leftarrow \Lambda(B_i^{\pi_{i,j}})$, \ 
	$r \leftarrow r + rstep$, \ and let \ 
	$i \leftarrow i+1$,
	\item $\rightarrow 2$ \  if $(i \le p)$ \ and desired bound has not been achieved, {\bf else} end.
\end{enumerate}
In tables below, we compare the length of the shortest vector by using the hill climbing reduction approach on the ideal lattice of rank 68 and the ideal lattice of rank 108.  
The value at HC-r represents the length of the shortest vector due to the hill climbing algorithm of radius $r$ in the corresponding stage. The value at HC-PSL2 represents the length of the shortest vector due to the hill climbing algorithm using PSL2 group elements. The LLL-RR (arbitrary precision floating point) version of LLL algorithm from the NTL library is used to reduce the lattice basis. The sample size taken in each step is 100.
\vs{2mm}

Initially, the length of the shortest vector obtained for the lattice of rank 68 after an application of the  LLL-RR algorithm with reduction parameter $\alpha =0.9$ was  4664. The shortest vector in the experiment in the table below corresponds to the reduction parameter $\alpha =0.9999$. We have compared the hill climbing reduction with different radii using the same parameter $\alpha =0.9999$. The length of the shortest vector was 2222 that corresponds to BKZ-10.

\begin{table}[H]
	\label{tab45}
	\centering
	\begin{tabular}{|c|c|c|c|c|c|c|c|c|c|}
		\hline
		Step &LLL & BKZ-10& HC-30 &HC-42& HC-48 & HC-54 &HC-66  & HC-PSL2\\
		\hline
		1 & 2890 & 2222 & 2172 & 2218& 2172 & 2189& 2154  &2255\\
		\hline
		5 & -- &--& 2152 & 2172 & 2172& 2172 & {\bf 2141} & 2182\\
		\hline
	\end{tabular}
	\vs{3mm} \caption{Comparison for an ideal lattice of rank 68}
\end{table}
Next, the length of the shortest vector obtained for the lattice of rank 108 after the application of the LLL-RR algorithm with reduction parameter $\alpha =0.9$  was  12,308. Since the shortest vector  corresponds to the LLL paramerer $\alpha =0.99999$, we compare the hill climbing reduction with different radii with the same parameter $\alpha =0.99999$.

\begin{table}[H]
	\label{tab46}
	\centering
	\begin{tabular}{|c|c|c|c|c|c|c|c|c|c|}
		\hline
		Step &LLL & BKZ-35& HC-50 &HC-60 & HC-70 &HC-80  & HC-100 &HC-PSL2\\
		\hline
		1 & 6275 & 4616&4258 & 4129& 4309 & {\bf 3835} & 4202 &4513\\
		\hline
		4 & -- &--& 4015  & 3938 & 4010 & 3835  & 4104 & 4337\\
		\hline
		7 & -- &--& 4015  & 3880 & 4010 & 3835  & 4006 & 4064\\
		\hline
	\end{tabular}
	\vs{3mm}\caption{Comparison for an ideal lattice of rank 108}
\end{table}
From above tables, we observe that the hill climbing algorithm with any radius produces smaller $l$  values than the LLL or BKZ algorithms. Among all values, HC-66 in table (\ref{tab45}) and HC-80 in table
(\ref{tab46}) have better approximations for the shortest lattice  vectors. 

In the next section we observe the hill climbing lattice basis reduction approach based on the radius of permutations. 
We observed a number of experimental results of lattice basis reduction using our hill climbing approach on bases each of which are already reduced by a single LLL operation.
In this section, we observe convergence patterns as follows:  

\begin{itemize} 
	
	\item[(i)] for the length of the {\it shortest lattice vector at each hillstep} versus  {\it radius}.  
	\item[(ii)] {\it average length and the shortest vectors at all hill steps} versus  
	{\it radius}.
	\item[(iii)] {\it average length} and {\it shortest length of all radii} versus {\it each hill step}.
	
\end{itemize}

For each step and each possible radius, we have taken a basis sample of 100 permutations.
Based on our computations, we conclude that the run-time to get the shortest lattice vector using hill climbing with variable radii is {\bf less} than the approach with fixed radius. Scatterplot with a smooth fitted lines of our experiments are given below.

\begin{figure}[H]
	\centering
	\includegraphics[width=\textwidth]{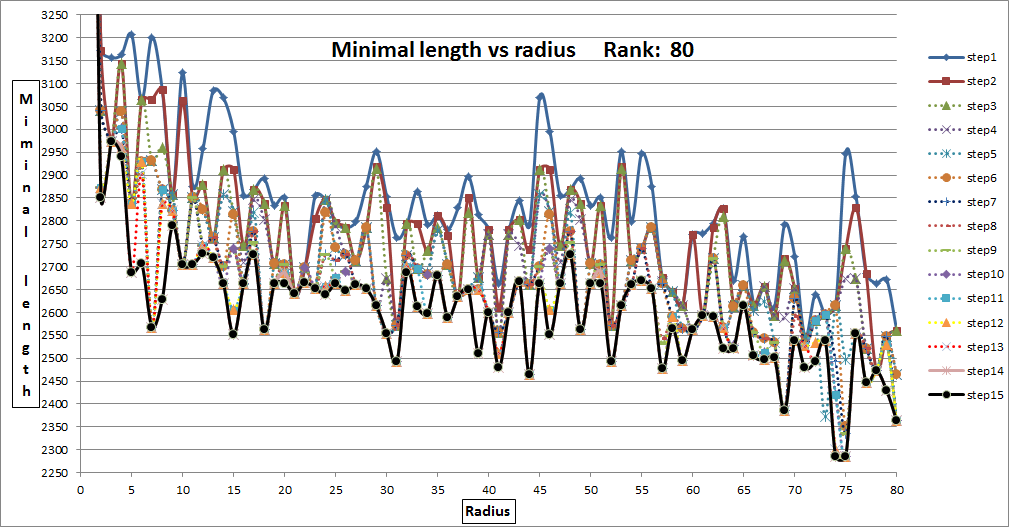}
	\vs{-4mm} \caption{Minimal length vs radius}
\end{figure}

\begin{figure}[H]
	\centering
	\includegraphics[width=\textwidth]{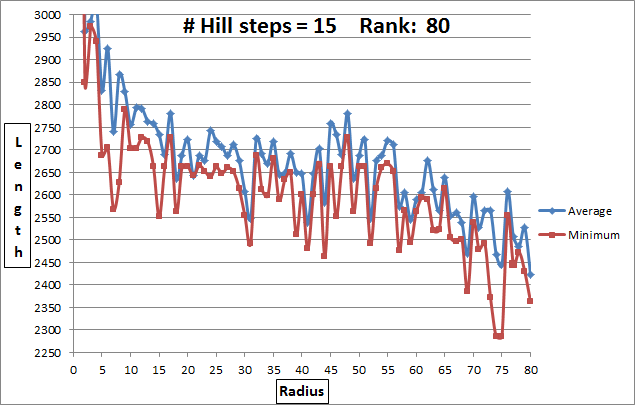}
	\vs{-4mm} \caption{Average length and the shortest length vs radius}		
\end{figure}
\begin{figure}[H]
	\centering
	\includegraphics[width=.72\textwidth]{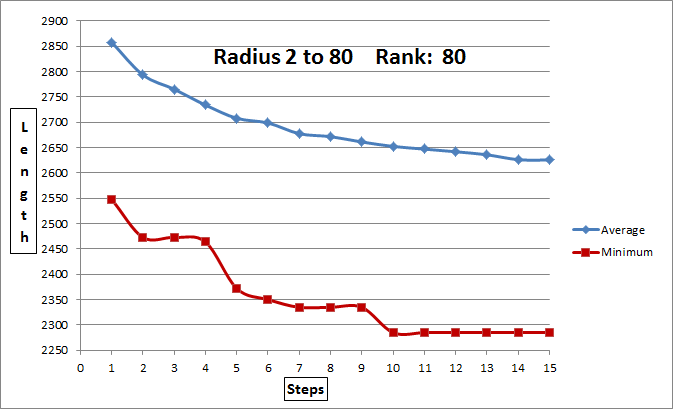}
	\vs{-4mm} \caption{Average length and the shortest length vs hill step}
\end{figure}
\normalsize

\section{Lattice diffusion and sublattice fusion}
Sending a basis to parallel nodes is not a good idea for a lattice basis that 
requires long runtime. 
Instead, we ``{\it diffuse}'' the given input basis into smaller blocks (not necessarily disjoint) and send these blocks to the nodes of a parallel server machines with proper instructions.
This method not only reduces the runtime but also reduces the value of the largest entries in the basis. 
As we get the output back from the server machines to the master client machine, we combine them and rearrange the basis vectors with a {\bf right} permutation. Then, we again diffuse the basis gradually increasing the block size.  The process continues until a desired bound it obtained.
Using this approach, we successfully achieved the lattice vectors whose estimated euclidean norm were less than  the given bound on the 
{\bf Lattice hall of fame} \cite{halloffame}. 

\subsection{Lattice diffusion and sublattice fusion algorithm (LDSF)}

\begin{itemize}
	\item[] \textbf{Input:}  Basis $B$ of $\cL \subset \bR^{n}, \ \beta :$
	Block Size, \ and  $N, \ M :$ \ parameters.
	\item[] \textbf{Output:} A lattice vector of ``minimal'' length close to $b$.
	\small
	\begin{enumerate}
		\item $m\leftarrow$ number of rows in $B\ \diamond \ \ l\leftarrow 1.$
		\item $i\leftarrow 1.$
		\item For $j=1\ \text{to}\ k$; \ $k$  is the number of server.
		\item $B_{j} \leftarrow$ a block matrix  taking randomly $\beta$ rows from $B$.
		\item Send $B_{j}$ to server $j \ \diamond \ \hat{B_{j}} \leftarrow$ result from server {j}.
		\item $B_{i}\leftarrow (\cup \hat{B_{j}})^{\pi} \ \ \diamond$ \ go to step 3 if $M\leq i\leftarrow i+1.$
		\item $b^{*}\leftarrow \min\limits_{i=1}^M \{\min||b_{i}||:b_{i}\in B_i\}.$ 
		\item End if $b^{*}\leq b.$
		\item Go to the step 2 if $(N\geq l\leftarrow l+1)\land (1\leq k \leftarrow k-1).$
	\end{enumerate}
\end{itemize}	

\begin{figure}[H]
	\centering
	\label{varldsf}
	\includegraphics[width=.85\textwidth]{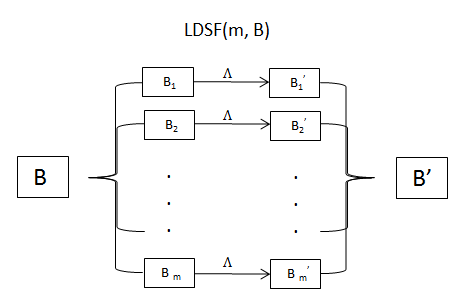}
	\vs{-6mm} \caption{LDSF algorithm}
\end{figure}

We presently consider a second operator which we denote by $C=\Sigma(m,n,B)$ which consists of  \ (i) \ applying a sample of $n$ permutations $\pi_1, \pi_2, \ldots , \pi_n$  to the input base $B$, \ (ii) \ computing $C_i= LDSF(m,B^{\pi_i})$, \ for $1\leq i \leq n$, and \ (iii) \ letting $C$ be the best basis among the resulting $C_i$. 

\begin{figure}[H]
	\centering
	\label{varldsf}
	\includegraphics[width=\textwidth]{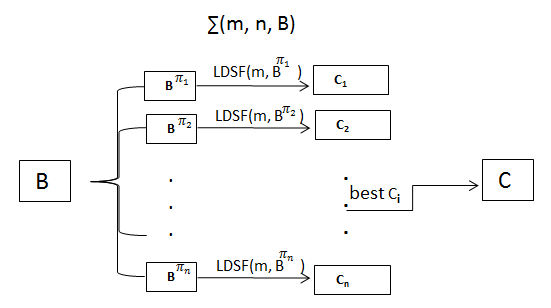}
	\vs{-6mm} \caption{Operator $\Sigma(m,n,B)$}
\end{figure}

If the time required to reduce a basis is very large, we use our multistage  lattice diffusion and sublattice fusion technique until we get significant changes of the basis entries. In fact, combination of lattice diffusion sublattice fusion algorithm with the hill climbing algorithm improves a lattice basis.
An explanation of multistage lattice diffusion and sublattice fusion plus a hill climbing algorithm is given below: 
\subsection{A 4-stage hybrid LDSF-HC process}
\label{hybrid}
In the first stage, given an initial  basis $B_0$, which is not reducible by LLL, 
we compute $LDSF(m, B_0)$ by breaking $B_0$ into $m$ blocks of sub-bases 
$B_{0,1}, B_{0,2}, \ldots , B_{0,m}$, and applying
the LLL-RR algorithm to each $B_{0,i}$, to obtain 
$\Lambda(B_{0,1}), \ldots , \Lambda(B_{0,m})$. We then fuse the $\Lambda(B_{0,i})$ to obtain  basis $B_1$ of $\cL$. This step helps to reduce the absolute value of the lattice basis entries. 
In the second stage, we obtain a random sample of $n$ right permutations $\pi_1 , \ldots , \pi_n$ and compute bases $C_j=LDSF(m,B_1^{\pi_j})$, for $1\leq j \leq n.$
Next, we select the best basis among the $\{C_j\}_{j=1}^n$ and call it $B_2$. The operations performed in stage 2 can be succinctly denoted by $B_2=\Sigma(m,n,B_1).$
Stage 3 is simply $B_3=\Sigma(l,n,B_2)$ where $l<m$.
Finally there is a termination stage computing the basis $B_4=\Lambda(B_3)$. Schematically, 

\begin{figure}[H]
	\centering
	\label{varldsf}
	\includegraphics[width=\textwidth]{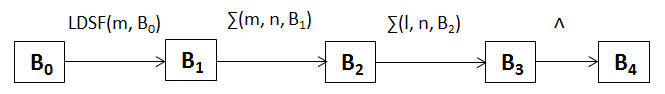}
	\vs{-6mm} \caption{4-stage hybrid LDSF-HC}
\end{figure}

If the lattice basis is not sufficiently reduced to use LLL in the 3rd stage, we add additional stages like stage 3 hoping for a better reduction. For example, in the experiments below, we use a  5-stage and 6-stage  models.

In  the experiments below, we compare our lattice diffusion and sublattice fusion algorithm with the fastest floating point version of LLL lattice reduction algorithm (LLL-FP) available in the NTL library with the parameter $\alpha=0.9$.

\subsection{Experiment} 
\label{hyb1}
The initial information is an ideal lattice of rank 300 obtained from \cite{halloffame}.

i)  The  length of each basis vectors is $\approx 10^{900}$.\\
ii) The initial time to reduce the lattice basis  with our device of 24 GB/3GHz i7  processor using the NTL version of LLL-RR, with reduction parameter $\alpha = 0.9$, is approximately a week. 
\begin{center}
	\begin{table}[H]
		\centering
		\begin{tabular}{|c|c|c|c|c|c|}
			
			\hline  
			\multicolumn{1}{|c|}{Stage}& \multicolumn{1}{c|}{\# Sample} & 		                   \multicolumn{1}{c|}{$\# Blocks$}&\multicolumn{1}{c|}{$llb$}&
			\multicolumn{1}{c|}{$lub$}&
			\multicolumn{1}{c|}{Runtime} \\
			\hline
			
			1& 1 & 3  & 19324936602 & 8659772341291 & 1hr 1 m \\
			\hline
			2 & 10  & 6 & 4509832663 & 62270235928 & 3 m 29 sec\\
			\hline
			3 & 5 & 3 & 2799147462  & 81977150923  & 33 m 51 sec   \\
			\hline
			4 & 5 & 2 & 397155  & 1249483 &  53 m 40 sec  \\
			\hline
			5 & 5 & 2  & 140973 & 236873&  1 hr  6 m \\
			\hline	
		\end{tabular}
		\caption{Reduction on an ideal lattice basis of order 300}
	\end{table}
\end{center}

{\bf Conclusion:} In  above experiment, the length of a shortest vector reduced from $\approx 10^{900}$ to 140973.

\end{document}